\documentclass[12pt,showpacs,aps]{revtex4}
\begin{document}

\preprint{}

\title{Magnetically induced vacuum decay }

\author{She-Sheng Xue}

\email{xue@icra.it}

\affiliation{ICRA, INFN and
Physics Department, University of Rome ``La Sapienza", 00185 Rome, Italy}



\begin{abstract}
We study the fermionic vacuum energy of vacua with and without being applied
an external magnetic field. The energetic difference of two 
vacua leads to the vacuum decaying and the vacuum-energy releasing. 
In the context of quantum field theories, we discuss why and how the 
vacuum energy can be released by spontaneous photon emissions and/or 
paramagnetically screening the external magnetic field. In addition, 
we quantitatively compute the vacuum energy released, the paramagnetic 
screening effect and the rate and spectrum of spontaneous photon emissions. 
The possibilities of experimentally detecting such an effect of vacuum-energy 
releasing and this effect accounting for the anormalous X-ray pulsar
are discussed. 
\end{abstract}

\pacs{12.20ds,
12.20fv }

\maketitle

\section{Introduction}\label{int}

The vacuum has a very rich physical content in the context of relativistic 
quantum field theories. It consists of extremely large number of virtual 
particles and anti-particles. The quantum-field fluctuations of the vacuum are 
creations and annihilations of these virtual particles and anti-particles 
in all possible energy-range. As a consequence of the quantum fluctuations
of bosonic and fermionic fields in the vacuum, the vacuum energy does not vanish. 
In quantum field theories for free and massless particles,   
the positive vacuum-energy (the zero-point energy) of virtual photons is given by,
\begin{equation}
{\cal E}_o=2\Big({V\over(2\pi)^3}\Big)\int d^3 p\epsilon(|p|),\hskip0.3cm
\epsilon(|p|)=\sqrt{p_x^2+p_y^2+p_z^2},
\label{eve}
\end{equation}
in a volume $V$ of the three-dimensional space, where the factor ``2'' is for 
polarization states. Analogously, with the {\it negative} 
and {\it non-degenerate} energy-spectrum of free virtual fermions, the fermionic 
vacuum energy is given by 
\begin{eqnarray}
{\cal E}_o &=& 4\Big({V\over(2\pi)^3}\Big)\int d^3 p|\epsilon_F(p)|,\label{ve}\\
\epsilon_F(|p|)&=& -\sqrt{p_x^2+p_y^2+p_z^2+m^2},
\label{fs}
\end{eqnarray} 
where $m$ is the mass of fermions and the factor ``4'' is for spin states.
In Eqs.(\ref{eve},\ref{ve}), the summation is over all possible momentum-states of quantum-field 
fluctuations. Up to the fundamental Planck scale $\Lambda_p$, 
the vacuum-energy Eqs.(\ref{eve},\ref{ve}) are constants, $|{\cal E}_o|\sim V\Lambda_p^4$.   

In the description of renormalizable and perturbative quantum field theories, the vacuum state as ground 
state is that all negative energy states are fully filled by the pairs of virtual fermions and 
anti-fermions. The virtual fermion in the negative energy state travels backwards in time indicating 
a virtual anti-fermion in the positive energy state travels forwards in time. The pairs of virtual 
fermions and anti-fermions are created by virtual photons and annihilated into virtual photons
in the time scale $\sim\hbar/mc^2$ and 
at the distance scale $\sim\hbar/mc$. While, real particles and antiparticles are excitation quanta 
upon the vacuum state. The vacuum energy Eqs.(\ref{eve},\ref{ve}) are 
dropped and set to be zero by the normal ordering of creation and annihilation 
operators, owing to the absolute value of the physical energy only determined up 
to a constant. The quantum-field fluctuations of the vacuum impacting on real particles 
and antiparticles are treated by the renormalization of quantum field theories. 
The descriptions of renormalizable 
and perturbative quantum field theories have been extremely successful, as examples, 
the Lamb-shift effect\cite{lamb} and electric charge renormalization. These effects
indeed exhibit the highly non-trivial structure of the quantum electromagnetic 
dynamics (QED) and its vacuum (ground) state. 

However, as shown by the Casimir effect\cite{casimir} that was experimentally 
evidenced\cite{exp}, the positive vacuum-energy (\ref{eve}) of virtual photons
is not just a trivial constant, when the quantum fluctuations of virtual photons 
of the vacuum state are confined within a finite volume by boundary conditions.
This effect shows that the vacuum state is modified by boundary conditions. 
From the energetical point of view, the Casimir effect can be physically understood 
as the following: (i) the continuous energy-spectrum (\ref{eve}) of electromagnetic 
fields is modified by boundary conditions to be discrete one; (ii) the vacuum energy of 
``final'' vacuum state, computed by the discrete 
energy-spectrum in a given finite volume $V$, is smaller than the vacuum energy of 
``initial'' vacuum state, computed by the continuous energy-spectrum (\ref{eve}) 
in the same volume; (iii) as a result, the vacuum gains energy and becomes energetically 
unstable and has to decay from the ``initial'' vacuum state to the ``final'' vacuum state 
by quantum-field fluctuations. The difference of vacuum energies between two vacuum states 
must be released and this leads to an attractive and macroscopic force observed 
in the Casimir effect.  

In this article, instead of modifying the energy-spectrum (\ref{eve}) of virtual photons
by boundary conditions, we attempt to study the variation of the vacuum-energy 
(\ref{ve}) by modifying the negative energy-spectrum (\ref{fs}) of virtual fermions 
by an external magnetic field. We try to find any possible observable effects of the 
vacuum decay due to such a modification, and discuss the possibilities that these effects could be 
experimentally tested. 
  
\section{Vacuum instability}\label{inst}

Within a space volume $V=L_x\cdot L_y\cdot L_z$, we introduce an external 
constant magnetic field $B$ along the z-axis. As well known as the Landau levels\cite{landau}, 
the negative energy-spectrum of virtual charged fermions is given by
\begin{equation}
\epsilon(p_z,n,\alpha)=-\sqrt{p_z^2+m^2+|e|B(2n+1)-eB\alpha},\hskip0.3cm n=0,1,2,3,
\cdot\cdot\cdot ,
\label{sh}
\end{equation}   
where $e$ and $\alpha=\pm 1$ are fermion's bare charge and helicity. This negative 
energy-spectrum is {\it degenerate} in the phase space of ($p_x,p_y$) and the degeneracy 
is $|e|BS/(2\pi)$, where the area $S=L_x\cdot L_y$. 

We define the vacuum state with $B=0$ as ``initial'' vacuum state and the vacuum state 
with $B\not=0$ as ``final'' vacuum state. The negative energy-spectrum (\ref{fs}) of 
``initial'' vacuum state is modified to the negative energy-spectrum (\ref{sh}) of 
``final'' vacuum state, due to the external magnetic field $B$. 
If the vacuum energy of the ``final'' vacuum state made by virtual fermions fully filling 
the negative energy-spectrum (\ref{sh}) is smaller than  that of the ``initial'' vacuum 
state made by virtual fermions fully filling the negative energy-spectrum (\ref{fs}), 
the vacuum state gains energy and must decay from the ``initial'' vacuum state to the ``final'' 
vacuum state by quantum-field  fluctuations. The difference of vacuum energies between two vacuum 
states must be released, leading to possibly observable effects. In order 
to verify this, we are bound to compute the energetic difference between two vacuum states 
respectively corresponding to $B=0$ and $B\not= 0$. 

The vacuum energy of the ``initial'' vacuum state $(B=0)$ is given by eq.(\ref{ve}).
Whereas, the vacuum energy of the ``final'' vacuum state $(B\not=0)$ is given by,    
\begin{equation}
{\cal E}_n=-\Big({|e|BS\over2\pi}\Big)\Big({ L_z\over2\pi}\Big)\int dp_z\sum_{n,\alpha}|
\epsilon(p_z,n,\alpha)|.
\label{nve}
\end{equation}
Both vacuum energies (\ref{ve}) and (\ref{nve}) are divergent up to the Planck scale 
$-V\cdot\Lambda_p^4$. By using the approaches of the dimensional 
regularization\cite{thooft} and $\xi$-function regularization\cite{db}, we compute 
the vacuum energies Eq.(\ref{ve}) for $B=0$ and Eq.(\ref{nve}) for $B\not=0$.
In Eq.(\ref{ve}), analytically continuing the dimension of the momentum integration from 
$3$ to $3+\epsilon$, where $\epsilon$ is a small complex parameter, we have
\begin{equation}
{\cal E}_o=\Big({V\pi\over(2\pi)^3}\Big)m^4\Gamma(-{\epsilon\over2}),
\label{ove1}
\end{equation}
where $\Gamma(x)$ is the Gamma-function. Analogously, in Eq.(\ref{nve}), analytically 
continuing the dimension of the momentum integration from $1$ to $1+\epsilon$, we have,
\begin{equation}
{\cal E}_n =-\Big({|e|BV\over4\pi^2}\Big)\Gamma(-{\epsilon\over2})\sum_{n,\alpha}
\Big[m^2+|e|B(2n+1)-eB\alpha\Big]^z,
\label{dnve}
\end{equation}
where $z=1+\epsilon/2$. Summing over helicity states $\alpha=\pm 1$ in Eq.(\ref{dnve}), 
we obtain 
\begin{eqnarray}
{\cal E}_n &=&-\Big({|e|BV\over4\pi^2}\Big)\Gamma(-{\epsilon\over2})\sum_n
\left[\Big(m^2+2|e|Bn\Big)^z+ \Big(m^2+2|e|B(n+1)\Big)^z\right],\nonumber\\
& = & -\Big({2|e|^2B^2V\over4\pi^2}\Big)\Gamma(-{\epsilon\over2})\left[\xi(-z,q)
+\xi(-z,q+1)\right],\hskip0.2cm q={m^2\over 2 |e|B},
\label{dnve1}
\end{eqnarray}
where $\xi(z,q)$-function is given by Eqs.(9.521), (9.531) and (9.627) in \cite{gr},
\begin{equation}
\xi(z,q)=\sum_{n=0}{1\over (n+q)^z},\hskip0.3cm\xi(-1,q)=-{B_2(q)\over2},\hskip0.3cm 
B_2(q)=q^2-q+{1\over6}.
\label{xi}
\end{equation}
The analytic continuation ``$z$'' has simply discarded the appropriate divergent 
terms and the continuation back to ``$z=1$'' ($\epsilon\rightarrow 0$) yields,
\begin{equation}
{\cal E}_n =V\Gamma(-{\epsilon\over2})\left[{|e|^2B^2\over3(4\pi^2)}+ {\pi 
(m^2)^2\over(2\pi)^3}\right].
\label{dnve2}
\end{equation}
These results Eq.(\ref{ove1}) for $B=0$ and Eq.(\ref{dnve2}) for $B\not=0$ are in agreement with 
Eqs.(3.8) and (4.2) in the ref.\cite{referee}. This provides us a consistent check of our computations.

As a result, the energetic difference of the ``initial'' vacuum state ($B=0$) and ``final'' 
vacuum state ($B\not=0$) is,
\begin{equation}
\Delta{\cal E}={\cal E}_n - {\cal E}_o =\Gamma(-{\epsilon\over2})
\left[{|e|^2B^2V\over3(4\pi^2)}\right].
\label{delta1}
\end{equation}
We find that in Eq.(\ref{dnve2}) the term depending on the fermion mass is 
completely canceled by Eq.(\ref{ove1}), as it should be. For $\epsilon\rightarrow 0$, 
the Gamma-function $\Gamma(-\epsilon/2)=-(2/\epsilon + {\rm const.})$, where the 
constant is an uninteresting combination of $\pi,\gamma$(Euler constant), etc. On 
the basis of the charge renormalization of the QED, we renormalize the charge 
$|e_r|=\sqrt{Z_3}|e|$ by the renormalization constant $Z_3=(2/\epsilon + 
{\rm const.})$ and we obtain,
\begin{equation}
\Delta{\cal E}=-\sum_f(Q_f^2){e_r^2B^2V\over3(4\pi^2)}=-8{\alpha \over3\pi}B^2V, 
\hskip0.5cm \alpha={e_r^2\over4\pi}={1\over137},
\label{delta2}
\end{equation}
where $\sum_f(Q_f^2)=8$ for all charged fermions in the standard model of particle 
physics. Eq.(\ref{delta2}) is about one percent of the total energy deposited by 
the external magnetic field $B$.

The energetic difference Eq.(\ref{delta2}) between the vacuum states ($B=0$)
and ($B\not=0$) is negative, indicating that the vacuum energy Eq.(\ref{ove1}) 
($B\not=0$) is smaller than the vacuum energy Eq.(\ref{dnve2}) ($B=0$), i.e., 
the vacuum state gains energy when the external magnetic field 
is applied upon it. The reasons are following. (i) In a finite volume $V$ and the finite 
momentum-cutoff at the Planck scale $\Lambda_p$, the total number of fermion states in the 
vacua of negative energy-spectra (\ref{fs}) and (\ref{sh}) are finite and all 
these fermion states of negative energy levels from $-\Lambda_p$ to $-mc^2$ are 
fully filled. (ii) The negative energy-spectrum (\ref{fs}) is not degenerate, 
while the negative energy-spectrum (\ref{sh}) is degenerate, and the total
numbers of fermion states of both cases are the same. (iii) On the basis of 
quantum-field fluctuations towards the lowest energy-state and the Pauli principle, 
when the external magnetic field $B$ is applied upon the vacuum, the vacuum 
reorganizes itself by fully filling all fermion states according to the 
degenerate negative energy-spectrum (\ref{sh}), instead of the non-degenerate 
one (\ref{fs}). As a consequence, the vacuum makes its total energy lower. 
As an analogy, the vacuum with the negative energy-spectrum (\ref{fs}) can be 
described as if a $N$-floors building, two rooms each floor and all rooms 
occupied by guests; while the vacuum with the negative energy-spectrum (\ref{sh}) 
a $M$-floors $(M<N)$ building, $2N/M$ rooms each floor and all rooms occupied by 
guests. The total numbers of rooms of two buildings are the same. Due to an 
external force, the $N$-floors building collapses to the $M$-floors building and 
the ``potential energy'' is reduced. 

In principle, due to the vacuum state gains energy when the external magnetic field 
is applied upon it, the vacuum becomes unstable and must decay and release the 
energy $\Delta{\cal E}$ (\ref{delta2}) by quantum-field fluctuations, analogously
to the dynamics for the Casimir effect, discussed in the section of introduction.  
If the vacuum state decays and the vacuum energy (\ref{delta2}) is released,
the phenomena and effects that could occur are following. (i) The vacuum acts 
as a paramagnetic medium that effectively screens the strength of the external 
magnetic field $B$ to a smaller value $B'<B$ for the total energy-density being,
\begin{equation}
{1\over2}B'^2={1\over2}B^2-8{\alpha \over3\pi}B^2;\hskip0.5cm B'=B\sqrt{1-
{16\over3\pi}\alpha}.
\label{hed}
\end{equation}
This phenomenon of paramagnetic screening could be possibly checked by appropriate 
experiments of precisely measuring the magnetic field strength. (ii) The 
vacuum-energy fluctuations could lead to the emission of neutrino and anti-neutrino 
pairs from the vacuum, since they are almost massless. This however is almost 
impossible for an experimental test.
(iii) Photons are spontaneously emitted analogously to the spontaneous photon 
emissions for electrons at high-energy levels decaying to low-energy levels in 
the atomic physics. This phenomenon should be possibly detected if any photons 
are emitted when the magnetic field $B$ is turned on.

In practice, it must be very complicate to measure this vacuum energy $\Delta{\cal E}$ 
(\ref{delta2}) releasing. In the following section, we discuss the rate and spectrum 
of spontaneous photon emissions.

\section{The rate and spectrum of spontaneous photon emissions}\label{emmision}

Let us assume the magnetic field $B=B(t)$ is adiabatically turned on
\begin{equation}
B(t)=\Big\{\matrix{0,& t= t^-\rightarrow-\infty\cr B, & t= t^+\rightarrow+\infty}
\label{bt}
\end{equation}
in the time interval $\Delta\tau=t^+-t^-$. Based on this assumption, we compute 
rate and spectrum of spontaneous photon emissions, purely due to the variation of 
vacuum energy (\ref{delta2}).

In order to obtain the rate and spectrum of spontaneous photon emissions, we need to compute 
a transition amplitude from the ``initial'' fermionic vacuum state ($B=0$) to the ``final'' fermionic 
vacuum state ($B\not=0$), i.e., from initial negative energy-states $\psi_i^{(-)}$ of fermions 
when $B=0$ to final negative energy-states $\psi_f^{(-)}$ of fermions when $B\not=0$. 
For simplifying computations of the transition amplitude at tree-level, we consider that 
the initial negative energy-states $\psi_i^{(-)}$ of fermions are zero-momentum states 
(${\vec p}_i=0$), and the final negative energy-states $\psi_f^{(-)}$ of fermions are all 
possible states. Beside, we chose the magnetic field $B$ is in $\hat z$-direction and the 
electric charge $e$ is renormalized. Thus, the initial negative energy-states $\psi_i^{(-)}
$ for $B=0$ and final negative energy-states $\psi_f^{(-)}$ for $B\not=0$ are given by    
\begin{eqnarray}
\psi_i^{(-)}&=&\left({1\over V}{m\over E_i}\right)^{1\over2}e^{iE_it}
\left(\matrix{0\cr\chi^\alpha}\right);\label{initial}\\
 \psi_f^{(-)}
& =& \left({1\over V}{m\over E_f}\right)^{1\over2}c_ne^{-{\xi^2\over2}}
H_n(\xi)e^{iE_ft-p_y^fy-p_z^fz}\left(\matrix{0\cr\chi^\alpha}\right)\label{final}\\
E_i&=&m,\label{ie}\\
E_f &=& \sqrt{m^2+(p^f_z)^2+eB(2n+1-\alpha)},
\label{ifstates}
\end{eqnarray}
where the spinor $\chi^\alpha$: $\sigma_z\chi^\alpha=\alpha\chi^\alpha$ for 
the helicity $\alpha=\pm 1$, $\xi=\sqrt{eB}(x-{p^f_y\over eB})$, $H_n(\xi)$ 
is the Hermite polynomial and 
$c_n=1/(2^{n\over2}\sqrt{n!}\pi^{1\over2})$. These negative energy
solutions can be obtained by the charge conjugation of corresponding positive 
energy-solutions. The probability of spontaneous photon emissions is 
related to the amplitude 
$|\epsilon_\mu^\beta J^\mu(k)|^2$\cite{iz}, where $\epsilon_\mu^\beta$ 
is the transverse polarization vector of photons emitted and 
\begin{equation}
J^\mu(k)=e\int d^4xe^{-ikx}\bar\psi_f^{(-)}\gamma^\mu\psi_i^{(-)},
\label{j1}
\end{equation}
where ``$k$'' is photon's energy-momentum. Using $\psi_i^{(-)}$ (\ref{initial}) 
and $\psi_f^{(-)}$ (\ref{final}), we integrate variables $t,y$ and $z$ in 
Eq.(\ref{j1}), which gives rise to $\delta$-functions for energy-momentum conservations. 
Armed with Eq.(7.376) in \cite{gr}, we integrate variable $x$ in Eq.(\ref{j1}),
\begin{equation}
\int dxe^{-ik_xx}e^{-{\xi^2\over2}}H_n(\xi)=(-i)^n({2\pi\over eB})^{1\over2}
e^{-ik_x{p^f_y\over eB}}H_n({k_x\over \sqrt{eB}})e^{-{k_x^2\over 2eB}}.
\label{j}
\end{equation}
The computation of the amplitude $|\epsilon_\mu^\beta J^\mu(k)|^2$ is 
straightforward in the spinor space. Taking average over helicities of 
initial states, summing over all final states $p^f_z,p^f_y$ and $n$ with 
degeneracy $|e|BS/(2\pi)$,
as well as the polarizations of photons emitted, we obtain,
\begin{eqnarray}
{1\over2}\sum_\beta|\epsilon_\mu^\beta J^\mu(k)|^2 &=& e^2e^{-{k_x^2\over eB}}
\sum_{n=1}^\infty{1\over2^nn!\pi}H^2_n({k_x\over \sqrt{eB}})(2\pi)
\delta[\omega_k-E^n_f+m]\left({m\over E^n_f}\right)\Delta\tau\nonumber\\
E^n_f&=&\sqrt{m^2+(k_z)^2+2eBn}
\label{re}
\end{eqnarray}
where $\omega_k=|k|$ is the photon energy and $\delta$-function for the 
energy-conservation. The term corresponding to $n=0$ has been dropped for 
energy-momentum conservations, since the $\delta$-function 
$\delta[\omega_k-E^{n=0}_f+m]$ only gives solution $|k|\equiv 0$. 
Because the problem is axial symmetric w.r.t. $\hat z$-direction, 
we can make substitutions $k^2_x\rightarrow k_\perp^2=k_x^2+k_y^2$, 
$k_x\rightarrow |k_\perp|$ and define $k_z=k_\parallel$ in Eq.(\ref{re}). 
The $\delta$-function in Eq.(\ref{re}) can be given as 
\begin{equation}
\delta[\omega_k-E^n_f+m]=\left({m+|k|\over eB}\right)\delta_{n,n_\circ},
\hskip0.2cm 
n_\circ={k_\perp^2+2|k|m\over 2eB},\hskip0.2cm E^{n_\circ}_f=m +|k|,
\label{n}
\end{equation}
where $n_\circ=1,2,3,\cdot\cdot\cdot$, indicating the energy-momentum of 
emitted photons is quantized. As a result, Eq.(\ref{re}) is,
\begin{equation}
{1\over2\Delta\tau}\sum_\beta|\epsilon_\mu^\beta J^\mu(k)|^2 = 
{e^2\over2^{n_\circ} n_\circ !\pi}e^{-{k_\perp^2\over eB}}H^2_{n_\circ}
({|k_\perp |\over \sqrt{eB}})(2\pi)\sum_f(Q^2_f)\left({m_f\over eB}\right),
\label{fre}
\end{equation}
where $\sum_f$ is over all flavors of charged fermions. We find that 
Eq.(\ref{fre}) does not explicitly depend on $k_\parallel$. For given 
$n_\circ=n_\circ(k_\perp,k_\parallel)$ in (\ref{n}), Eq.(\ref{fre}) 
(the probability of photon emissions) is very small for 
$|k_\parallel|\gg |k_\perp|$, because the polynomial 
$H^2_{n_\circ}({|k_\perp |\over \sqrt{eB}})$ in Eq.(\ref{fre}) is 
very small for small values of $|k_\perp|$. Thus, the most probability 
of spontaneous photon emissions occurs for $|k_\perp|\gg |k_\parallel|$, indicating 
most emitted photons are near in the plane perpendicular to the magnetic 
field $B$. This is analogous to the phenomenon of synchrotron radiation. 

The probability $p_{n_\gamma}$ corresponding to the emission of $n_\gamma$ 
photons, when neither the momentum nor the polarizations are observed, is 
given by the Poisson distribution\cite{iz},
\begin{equation}
p_{n_\gamma} = {\bar n_{\gamma}\over n_{\gamma}!}e^{-\bar n_{\gamma}},
\label{pn}
\end{equation}
where $\bar n_{\gamma}$ is defined by
\begin{equation}
\bar n_{\gamma} =\int d\tilde k{1\over2}\sum_\beta|\epsilon_\mu^\beta 
J^\mu(k)|^2,\hskip0.3cm \int d\tilde k\equiv \int{d^3k\over 
(2\pi)^32\omega_k}\simeq\int\left({dk_\parallel dk_\perp\over 
(2\pi)^22}\right)_{\omega_k\simeq |k_\perp|},
\label{np}
\end{equation}
which is actually the average number of emitted photons, 
$\bar n_{\gamma}=\sum_\circ^\infty n_{\gamma}p_{n_\gamma}$. 
The number- and energy-spectrum of spontaneous photon emissions in 
a phase space element $d\tilde k$ and a unit of time are given by
\begin{equation}
{d\bar n_{\gamma}\over d\tilde k} ={1\over2\Delta\tau}\sum_\beta|
\epsilon_\mu^\beta J^\mu(k)|^2,\hskip0.5cm 
{d\bar \epsilon_{\gamma}\over d\tilde k} ={1\over2\Delta\tau}
\omega_k\sum_\beta|\epsilon_\mu^\beta J^\mu(k)|^2,
\label{snp}
\end{equation} 
which are determined by Eq.(\ref{fre}).

We estimate that $\sqrt{eB}\simeq 0.244$eV for $B=10^5G$ achieved in 
the laboratory today\cite{magnet} and $\sqrt{eB}\simeq 24$KeV for 
$B\simeq 10^{15}G$ around neutron stars, i.e., $\sqrt{eB}\ll m_f$. 
We consider the limit for emitted photons whose energy-momentum 
$|k|\ll m_f$. From Eq.(\ref{n}), we have $n_\circ=|k|m_f/(eB)$, i.e.,
$|k|=\omega_k=n_\circ eB/m_f$, showing the energy of emitted photons 
is quantized in the unit $eB/m_f$. As seen from Eq.(\ref{fre}), the 
probability is exponentially suppressed for large values of $k_\perp^2/ eB$ 
and also suppressed by $1/(2^{n_\circ} n_\circ !)$ for large values of $n_\circ$. 
As a consequence, most photons emitted should have the momentum $
|k_\perp|\sim eB$ in the infrared region and quantized 
$|k_\perp|\simeq\omega_k=n_\circ eB/m_f$ for the small values of $n_\circ$.  
For $n_\circ=1, |k_\perp|\simeq |k|=\omega_k=eB/m_f$ and $H_1(x)=2x$, 
we have the rate, 
\begin{equation}
{1\over2\Delta\tau}\sum_\beta|\epsilon_\mu^\beta J^\mu(k)|^2=
4e^2e^{-{k_\perp^2\over eB}}k_\perp^2\sum_f(Q_f^2){m_f\over (eB)^2}.
\label{pre1}
\end{equation}
for the number- and energy- spectrum (\ref{snp}) of spontaneous photon emissions. 
This spectrum shows $k_\perp^2$-dependence in the low-energy region 
and $\exp(-{k_\perp^2\over eB})$-dependence in the high energy region, 
respectively similar to the Rayleigh-Jeans part and the Wien part of 
the spectrum of the black-body   
radiation. However, the energy-momentum $|k_\perp|\simeq \omega_k$ is 
quantized, in this sense, it is more analogous to the Wigner spectrum of 
the distribution of discrete energy-levels of atoms and nuclei. 

This is a preliminary study of the possibilities of releasing vacuum energy
$\Delta E$ (\ref{delta2}). In reality, the way of releasing the vacuum energy $\Delta E$
must be complicated and the effect of the back-reaction to the vacuum should be 
considered. These are subjects for future work.

\section{Possibly experimental test}\label{test}

Strictly speaking, the rate and spectrum of spontaneous photon emissions 
should depend on the way of turning on the magnetic field $B(t)$. 
The reason for us to make the assumption of adiabatical turning on magnetic field 
bases on the time-scale of quantum-field fluctuations of the vacuum is much 
smaller than $\Delta\tau$ (\ref{bt}). At this preliminary step, we regard the rate and 
spectrum of spontaneous photon emissions computed in the previous section
as a theoretical analysis in the adiabatical assumption. 

In practice, however, it must be difficult to measure the rate and spectrum 
of spontaneous photon emissions, purely attributed to the vacuum effect 
for the following reasons. (i) The variation of magnetic field with time must induce
a variation of electric field with time, as a result, soft photons are emitted. These
photons are hardly distinguished from photons spontaneously emitted due to the vacuum effect 
discussed in this article. (ii) The volume $(\Delta\tau\cdot c)^3$, in 
which vacuum states at different points of the space-time are 
causally-correlated, can be much larger than the volume $V$ in which the 
constant magnetic field is created. In addition, we know that the relaxation 
time of quantum-field fluctuations of vacuum states is of the 
order of $\hbar/m_ec^2\sim 10^{-20}$sec. With this very short relaxation 
time, all causally-correlated vacuum states at different points of the 
space-time can rapidly decay to the lower energy states before the maximum 
value $B$ of the external magnetic field is reached. In this case, we should not 
expect to detect spontaneous photon emissions only from the volume $V$, 
corresponding to the total vacuum-energy releasing given in Eq.(\ref{delta2}), 
since photons can be spontaneously emitted from the whole volume 
$\sim(\Delta\tau\cdot c)^3$ where vacuum states are causally-correlated. 

In the experiment of detecting the Casimir effect, two large parallel perfectly 
conducting plates of sizes $L^2$ at a distance $a$ ($L\gg a$) separate 
causal-correlation of the quantum-field fluctuations of virtual photons inside 
the volume $aL^2$ between two plates from the quantum-field fluctuations of virtual 
photons outside of two plates. Since the creations and annihilations of virtual 
fermions and anti-fermions are related to the annihilation and creation of 
virtual photons, we assume the causal-correlation between vacuum states at 
two different points of the space-time is mediated by virtual photons. 
Inspired by the experiment for detecting the Casimir effect, we should use 
perfectly conducting box to isolate the volume $V$, where the external magnetic 
field is adiabatically turned on, from the space-time outside the box. 
The photon monitor and instrument of measuring magnetic field strength should be 
properly installed inside the box to detect the possible effect of spontaneous photon 
emissions. It is obvious that much sophisticate experiments must be proposed 
and/or another brilliant ideas for experiments have to be created, 
in order to detect such effects.  

Though such vacuum effects are difficult to be measured in a ground Laboratory, they 
could be probably observed in astrophysics events of photon and neutrino emissions, 
due to a very large variation of 
strength of magnetic fields in astrophysics processes (e.g., supernova explosions and 
neutron stars). Given the size of a neutron star of the order $10^6$cm and the strength of 
magnetic fields of the order of $10^{13}-10^{15}G$, we can estimate the maximum 
total vacuum-energy releasing is of the order of $10^{42}-10^{46}$erg from 
Eq.(\ref{delta2}). It is worthwhile to mention that such vacuum effects could account for 
the anomalous X-ray pulsar\cite{xray}.

Rotating Kerr black holes also can possess a large variation of the strength of 
magnetic fields, the vacuum effects discussed in this article can take place around 
the black hole. We do not know that any phenomenon relating to these vacuum effects 
can be observable. Regarding the applications of these vacuum effects to astrophysical 
processes of neutron stars and black holes, we have to consider whether or not the 
effect of gravitational field should be taken into account. If the gravitational field is 
significantly strong enough, the vacuum effects should be modified, see for example 
\cite{geyer}. In the neutron star case of size $10^6$cm, mass one solar mass and magnetic 
field $10^{13}$G, we can estimate that the force of gravitational field acting on 
an electron is about $2\cdot 10^{-12}$ Newton, while the force of magnetic 
field acting on an electron is about $5$ Newton. This implies that the gravitational 
field effects on the vacuum effects can be negligible in the neutron star cases.
However, in the Kerr black hole case, the effects of gravirational field on the black hole's 
horizon must not be negligible. It is worthwhile to study such gravitational effects.   
   
\section{Conclusion}\label{con}
Before ending this article, we wish to clarify that the effect presented here is essentially 
different from the well-known Heisenberg-Euler-Schwinger process\cite{ehs} 
on the view point of dynamics, though both of them are related to an external 
electromagnetic field. Let us first recall the main points of the 
Heisenberg-Euler-Schwinger process.

The quantum-field fluctuations of virtual particles and antiparticles are rather 
perturbative, for their space and time variations being much smaller than $\hbar/mc$ 
and $\hbar/mc^2$. The vacuum is vary stable against such small quantum-field fluctuations 
of virtual electrons and positrons (and other fermions), for the reasons: (i) 
all negative energy-states are fully filled; (ii) the energy-gap $2m$ is the potential barrier 
to block small quantum-field fluctuations leading to a vacuum instability. The probability of 
the vacuum breakdown and decaying by the pair-creation of real electrons and positrons 
(the probability of tunneling the barrier) is exponentially suppressed, practically is 
zero. The external electric field applied
upon the vacuum effectively reduces the energy-gap of barrier and the vacuum is 
polarized along the direction the electric field. 
When the electric field is increased to the critical field $E_c\simeq m^2c^3/e\hbar$, the 
probability of virtual electrons and positron tunneling the barrier significantly 
increases, leading to the phenomenon of the vacuum breakdown
by pair-creation of electrons and positrons in the Heisenberg-Euler-Schwinger process. 

It is important to notice that the Heisenberg-Euler-Schwinger process is a purely electric
effect from the dynamics point of view: the electric field instead of magnetic field makes 
the vacuum decaying and pair-production be feasible. 
On the contrary, the effect presented here is purely magnetic. 
As discussed in sections (\ref{inst}-\ref{emmision}), the dynamics of 
inducing vacuum decaying is completely due to the fact that the external magnetic field 
modifies the negative energy-spectrum from non-degenerate one to degenerate one, as a
result shifts the negative energy of the vacuum downwards, 
rather than reduces the energy-gap $2m$ separating the negative energy-spectrum from
the positive energy-spectrum. In contrast, the electric field in the 
Heisenberg-Euler-Schwinger process does not change negative energy-spectrum of the vacuum.
The possible phenomena of the presented effect are screening magnetic
field and spontaneous photon emissions, instead of pair-production of electrons and 
positions. In principle, the presented effect can occur for any value of magnetic field 
and non critical value of field is required. This is very different from the critical electric field 
in the Heisenberg-Euler-Schwinger process.
In fact, the dynamics of this effect presented here is analogous to the dynamics of the Casimir 
effect, rather than the dynamics of the Heisenberg-Euler-Schwinger one. The situation 
somehow resembles the phenomenon of chiral gauge anomalies: chiral fermions are derived out
of the Dirac sea (vacuum) by a chiral gauge field, leading to anomalous particles production.

In summary, we propose an idea of releasing vacuum-energy from the 
vacuum by introducing an external magnetic field. This idea is originated from (i) 
the fermionic structure of the vacuum owing to quantum field 
theories and the Pauli principle; (ii) the external magnetic field modifying 
the fermion spectrum of the vacuum from non-degenerate one to degenerate 
one; (iii) the quantum-field fluctuations of virtual particles in the vacuum must 
lead the vacuum state to the lowest energy-state. We illustrate this idea by giving 
an explicit computation of vacuum energy and showing the vacuum-energy releasing 
is about one percent of the total energy stored in the external magnetic field. 
We point out that such a vacuum-energy releasing could be realized by paramagnetic 
screening and/or spontaneous photon emissions. We compute the possible rate and spectrum of 
spontaneous photon emissions. 
In addition, we discuss the difficulties to observe such vacuum effects and propose a 
possible experiment to detect these effects.    
In today's laboratory, the magnetic field strength has been reached up to 
(greater than) $10^5$G and large stored energy up to (larger than) tens or 
hundreds of MJ \cite{magnet}. We expect a 
sophisticate experiment in near future to verify the phenomena and effects 
of the vacuum-energy releasing via paramagnetic screening and spontaneous 
photon emissions induced by external magnetic fields. This is important 
for the understanding of the fermion structure of the vacuum of
quantum field theories and any possibly prospective applications.





\end{document}